\documentclass[twoside,twocolumn,9pt]{article}

\usepackage{extsizes}
\usepackage[super,sort&compress,comma]{natbib} 
\usepackage[version=3]{mhchem}
\usepackage[left=1.5cm, right=1.5cm, top=1.785cm, bottom=2.0cm]{geometry}
\usepackage{balance}
\usepackage{mathptmx}
\usepackage{sectsty}
\usepackage{graphicx} 
\usepackage{lastpage}
\usepackage[format=plain,justification=justified,singlelinecheck=false,font={stretch=1.125,small},labelfont=bf,labelsep=space]{caption}
\usepackage{float}
\usepackage{fancyhdr}
\usepackage{fnpos}
\usepackage[english]{babel}
\addto{\captionsenglish}{%
  
}
\usepackage{array}
\usepackage{droidsans}
\usepackage{charter}
\usepackage[T1]{fontenc}
\usepackage[usenames,dvipsnames]{xcolor}
\usepackage{setspace}
\usepackage[compact]{titlesec}
\usepackage{hyperref}

\usepackage{epstopdf}

\usepackage{multirow}

\usepackage{chemformula} 
\usepackage{chemscheme}

\definecolor{cream}{RGB}{222,217,201}

\begin{document}

\pagestyle{fancy}
\thispagestyle{plain}
\fancypagestyle{plain}{

\renewcommand{\headrulewidth}{0pt}
}

\makeFNbottom
\makeatletter
\renewcommand\LARGE{\@setfontsize\LARGE{15pt}{17}}
\renewcommand\Large{\@setfontsize\Large{12pt}{14}}
\renewcommand\large{\@setfontsize\large{10pt}{12}}
\renewcommand\footnotesize{\@setfontsize\footnotesize{7pt}{10}}
\makeatother

\renewcommand{\thefootnote}{\fnsymbol{footnote}}
\renewcommand\footnoterule{\vspace*{1pt}%
\color{cream}\hrule width 3.5in height 0.4pt \color{black}\vspace*{5pt}} 
\setcounter{secnumdepth}{5}

\makeatletter 
\renewcommand\@biblabel[1]{#1}            
\renewcommand\@makefntext[1]%
{\noindent\makebox[0pt][r]{\@thefnmark\,}#1}
\makeatother 
\renewcommand{\figurename}{\small{Fig.}~}
\sectionfont{\sffamily\Large}
\subsectionfont{\normalsize}
\subsubsectionfont{\bf}
\setstretch{1.125} 
\setlength{\skip\footins}{0.8cm}
\setlength{\footnotesep}{0.25cm}
\setlength{\jot}{10pt}
\titlespacing*{\section}{0pt}{4pt}{4pt}
\titlespacing*{\subsection}{0pt}{15pt}{1pt}

\fancyfoot{}
\fancyfoot[LO,RE]{\vspace{-7.1pt}\includegraphics[height=9pt]{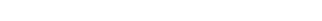}}
\fancyfoot[CO]{\vspace{-7.1pt}\hspace{13.2cm}\includegraphics{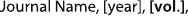}}
\fancyfoot[CE]{\vspace{-7.2pt}\hspace{-14.2cm}\includegraphics{head_foot/RF}}
\fancyfoot[RO]{\footnotesize{\sffamily{1--\pageref{LastPage} ~\textbar  \hspace{2pt}\thepage}}}
\fancyfoot[LE]{\footnotesize{\sffamily{\thepage~\textbar\hspace{3.45cm} 1--\pageref{LastPage}}}}
\fancyhead{}
\renewcommand{\headrulewidth}{0pt} 
\renewcommand{\footrulewidth}{0pt}
\setlength{\arrayrulewidth}{1pt}
\setlength{\columnsep}{6.5mm}
\setlength\bibsep{1pt}

\makeatletter 
\newlength{\figrulesep} 
\setlength{\figrulesep}{0.5\textfloatsep} 

\newcommand{\topfigrule}{\vspace*{-1pt}%
\noindent{\color{cream}\rule[-\figrulesep]{\columnwidth}{1.5pt}} }

\newcommand{\botfigrule}{\vspace*{-2pt}%
\noindent{\color{cream}\rule[\figrulesep]{\columnwidth}{1.5pt}} }

\newcommand{\dblfigrule}{\vspace*{-1pt}%
\noindent{\color{cream}\rule[-\figrulesep]{\textwidth}{1.5pt}} }

\makeatother

\twocolumn[
  \begin{@twocolumnfalse}
{\includegraphics[height=30pt]{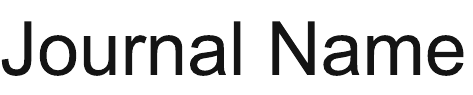}\hfill\raisebox{0pt}[0pt][0pt]{\includegraphics[height=55pt]{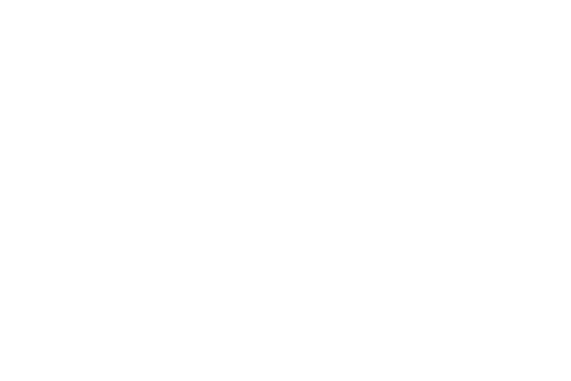}}\\[1ex]
\includegraphics[width=18.5cm]{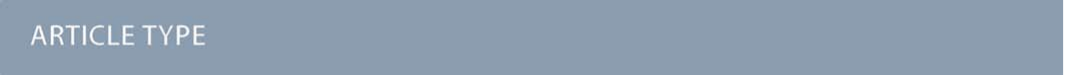}}\par
\vspace{1em}
\sffamily
\begin{tabular}{m{4.5cm} p{13.5cm} }

\includegraphics{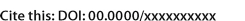} & \noindent\LARGE{\textbf{%
	Fast and accurate simulation of Raman spectra of gold-organic systems$^{\dag}$
}} \\
\vspace{0.3cm} & \vspace{0.3cm} \\

 & \noindent\large{%
	Auguste Tetenoire,$^{\ast}$\textit{$^{a,b}$}
	Vijaya Raghavan Kannan,\textit{$^{a,c}$}
	Mika\"el Kepenekian,$^{\ast}$\textit{$^{a}$}
	and Arnaud Fihey$^{\ast}$\textit{$^{a}$}
} \\

\includegraphics{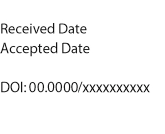} & \noindent\normalsize{%
Resolving the spectral Raman signature of molecules grafted on a metallic support is often a difficult task, in which quantum chemistry methods allow for precious additional rationalization and signal attributions, especially to probe the formation of a bond with the support. In the specific case of gold-organic architectures based on a Au-C bond, only a limited amount of experimental and theoretical reference data are available in the literature, and Raman simulations based on quantum mechanics quickly become unaffordable with the size of the system.
In this work, we evaluate the precision of a cost-efficient DFTB method to simulate Raman spectra of gold-organic systems at different scales, from gold complexes to functionalized gold surfaces. After a validation of the method through a careful comparison of DFTB Raman spectra of organometallic gold(I) and (III) complexes to DFT and experimental reference data, we discuss the case of molecules grafted on gold aggregates. For these simulations, the choice of the model (cluster or periodic surface) appears to be critical, and significant differences arise (positions and intensities of the peaks) when considering a full metallic slab, as allowed by the low computational cost of the method.
} \\

\end{tabular}

 \end{@twocolumnfalse} \vspace{0.6cm}

  ]

\renewcommand*\rmdefault{bch}\normalfont\upshape
\rmfamily
\section*{}
\vspace{-1cm}

\footnotetext{\textit{%
$^{a}$~Univ Rennes, ENSCR, CNRS, ISCR (Institut des Sciences Chimiques de Rennes) – UMR 6226, F-35000 Rennes, France;
$^{b}$~Laboratory of Computational Chemistry and Biochemistry, École Polytechnique Fédérale de Lausanne (EPFL), Lausanne 1015, Switzerland;
$^{c}$~Univ Rouen Normandie, INSA Rouen Normandie, CNRS, Normandie Univ, Institut CARMeN UMR 6064, INC3M FR 3038, F-76000, Rouen, France.
E-mail: mikael.kepenekian@univ-rennes.fr;
E-mail: arnaud.fihey@univ-rennes.fr};
E-mail: auguste.tetenoire@epfl.ch}

\footnotetext{{\dag}~Electronic Supplementary Information (ESI) available:
See DOI: 00.0000/00000000.}

\section{Introduction}

Raman spectroscopy is a cornerstone technique for non-destructive surface characterization, enabling the identification of molecular bonding and structural properties at interfaces.
Its enhanced variants, such as surface-enhanced Raman spectroscopy (SERS),~\cite{Ferrari2002,Nibbering2005,Wshipp2017} tip-enhanced Raman spectroscopy (TERS),~\cite{Lleru2012,Graham2016,Dick2016,Bzrimsek2017,Kusch2018,Perez-jimenez2020} and scanning near-field optical microscopy (SNOM),~\cite{Hecht2000,Hayazawa2002,Zhang2017} have revolutionized the study of adsorbates on plasmonic substrates like gold. 
However, simulating these techniques remains computationally prohibitive, particularly for large-scale systems, due to the intricate coupling of electronic, vibrational, and plasmonic-field effects. 
The variety of adsorbed molecules on gold surface is large, but thiol-based (Au-S) surface modification~\cite{Ramasamy2023} represent a large majority and have been already extensively studied with Raman spectroscopy experiment~\cite{Virdee1988,Szafranski1998,Loreen2004,Orendorff2005,Jing2007,Tripathi2013,Holze2015,Sun2017,Yang2017,Zhang2018,Nyamekye2021} and simulation.~\cite{Zhao2014,Peng2021,Merlen2022}
In contrast, gold-carbon (Au-C) bonds, despite their emerging relevance in catalysis and nanotechnology, remain understudied in both experimental and theoretical Raman spectroscopy. 

Theoretical spectroscopic investigations of Au-C materials face significant hurdles, and most of the first calculations were considering the vibrational intensities of small clusters only.~\cite{Liu2013,Zhao2004}
Density functional theory (DFT), while successful in modeling \ch{Au20} gold clusters and their vibrational spectra,~\cite{Ahmad2014,Berisha2018,Mohamed2018,Luo2020,Wang2019,Merlen2015} struggles with extended systems such as gold surfaces due to prohibitive computational costs. 
Existing experimental studies on Au-C vibrations are sparse, with pre-2000 experimental work limited to basic complexes~\cite{Shaw1973,Rice1975,Patterson1985,Feilchenfeld1989,Gao1990,Bruni1993} and post-2000 efforts only marginally expanding this scope.~\cite{Mrozek2001,Maity2013,Tijunelyte2017}
This gap underscores the need for robust computational tools capable of addressing both accuracy and scalability in studying Au-C bonding.

In this context Density Functional Tight-Binding (DFTB) offers a promising alternative,~\cite{Elstner1998} as an approximated and parameterized DFT method that balances computational efficiency with the predictive power and versatility of its parent framework. 
Unlike machine learning approaches, which are primarily deployed for spectral analysis~\cite{Liu2017,Sevetlidis2019,Weng2020,Kazemzadeh2022,Qi2023} and remain nascent for direct Raman simulations,~\cite{Hu2019,Msommers2020,Kapil2023,Berger2024,Xu2024}  DFTB methodologies have already been applied for such spectroscopy,~\cite{Witek2004,Ashtari-jafari2022}
and enable full quantum chemical simulations of large-scale structures, including metallic surfaces, at a fraction of the cost of DFT. 

Since the first effort of DFTB parametrization for gold-organic interactions, conducted in the second-order self-consistent charges framework (SCC-DFTB2) and mainly focused on describing thiolate organic moieties on gold slabs (\textit{auorg} set of parameters),~\cite{Fihey2015a} several DFTB studies have been dedicated to describe gold-thiolates architectures at the nanoscale. For instance, the structures and optical properties of functionalized gold nanoclusters have been shown to be accurately reproduced compared to DFT and experimental references,~\cite{GNC-transfer,GNC-DFTB-2023} and this parametrization was also used to model electron transport in molecular junctions.~\cite{DFTB-transport} So far, the assessment of the DFTB ability to describe Au-C based systems is lacking. In a recent work, we conducted a first example of a successful DFTB rationalization of the structure and Raman signal of calixarene-coated gold NPs, allowing to rationalize the nature of the interactions at the gold-organic interface.~\cite{Tetenoire2024} 

In this work, we aim at generalizing this approach and assess the validity of DFTB for simulating Raman spectra of Au-C based systems. We dedicate the first section of this work to gold-organic complexes, comparing its accuracy against DFT  and experimental data when available. 
We then demonstrate its unique capability to model extended systems—such as gold surfaces—that are not easily tractable with conventional DFT. 
Our analysis reveals that DFTB not only reproduces key Au-C stretching and ligand deformation modes, but also captures substrate-dependent frequency shifts and intensity variations. 
By bridging the gap between cluster-scale and surface-scale simulations, this approach unlocks new avenues for studying interfacial bonding in realistic environments. 


\section{Methods}

\subsection{SCC-DFTB computational details}

Self-Consistent Charges Density Functional Tight-Binding (SCC-DFTB) calculations were conducted as implemented in the version 22.2 of the DFTB+ program,~\cite{Hourahine2020} using the \textit{auorg} set optimized for the description of gold-organics interactions,~\cite{Fihey2015a} and its extension including phosphorous.~\cite{Vuong2020}
Empirical dispersions were included using the Grimme-D3 correction~\cite{Grimme2011} with Becke-Johnson damping parameters of $a_{1}$=0.5719, $a_{2}$=3.6017, $s_{6}$=1.0000, $s_{8}$=0.5883. The electronic ground state was determined with a SCC convergence threshold of 10$^{-5}$~Ha.
Geometry optimizations were performed using a force criterion of 10$^{-5}$~Ha/Bohr,
while for Hessian calculation the force criterion was set to 10$^{-8}$ Ha/Bohr.

The gold bulk was described in periodic boundary conditions using a $\Gamma$-centered 20$\times$20$\times$20 Monkhorst-Pack k-point grid to sample the Brillouin zone.~\cite{Monkhorst1976} Upon complete relaxation, a lattice parameter of 4.05~{\AA} is obtained, deviating only slightly (0.84~\%) from the experimental value of 4.08~{\AA}.~\cite{Ashcroft}
Au(111) surfaces were simulated using slabs consisting in 4 layers of a 20.0$\times$19.8~{\AA} supercell (ESI{\dag}) with a vacuum of 70.1~{\AA}. The electronic structure of the slab was described using $\Gamma$-point calculations.
Geometry optimizations were performed by applying no constraint on the three outermost layer of the slab and the adsorbed molecules, while the lattice vectors and the atomic positions of the bottom layers were kept frozen in order to emulate the gold bulk.

Subsequent computations of Raman spectra were obtained following the implementation in DFTB+ by Witek \textit{et al.} of the Placzek theory\cite{Witek2004} (see ESI{\dag} for additional details). For the gold complexes and functionalized Au$_{20}$, all atoms were free to move, while for the  functionalized surfaces we consider the displacement of all the atoms of the adsorbed molecule as well as the surface gold atoms bound to the molecule (see ESI{\dag} for a discussion of the impact of the number of frozen atoms in the gold slab).

\subsection{DFT computational details}
DFT reference molecular calculations for gold complexes were performed with the Amsterdam Density Functional (ADF) program, a part of the Amsterdam Molecular Simulation package (AMS 2020.101 release),~\cite{ADF2001} using the revPBE functional\cite{rev-PBE} with Grimme-D4 correction,~\cite{Caldeweyher2019} and atoms were described using the all-electron quadruple-$\zeta$ Slater-type orbitals basis set with four polarization functions (QZ4P).
In addition, we performed for comparison purposes the same calculations with a less costly PBE/DZP method, that resembles more closely the lighter DFT protocol used for the DFTB parameterization in \textit{auorgap}.\cite{Fihey2015a,Vuong2020} Both DFT methods results in very similar Raman spectra (see ESI{\dag}), with only modifications on the relative intensities. We thus discuss the DFTB/DFT comparison only on the basis of a revPBE/QZ4P method.

In both DFT methods, scalar relativistic effects were introduced \textit{via} the zeroth-order regular approximation (ZORA).~\cite{ZORA} Geometry relaxations were performed until forces were smaller than 5.10$^{-5}$~Ha/{\AA}. DFT Raman spectra were obtained by computing Raman intensities  within the AMS implementation (see ESI{\dag}).~\cite{vanGisbergen1996,vanGisbergen1999}

\section{Results and Discussions}

In the following sections, we first focus on the smallest gold-carbon architectures, \textit{i.e.} gold(I) and gold(III) complexes, for which both DFT and DFTB Raman calculations are applicable. 
They are compared, when available, with experimental data. 
We then describe small nano-objects (functionalized Au$_{20}$) that are commonly used as a cluster representation of the gold-organic interface, along with a fully periodic representation of the surface, and discuss the impact of the model on the computed Raman properties for a simple attached chemical group (\ch{CH3}). 
Finally, more complex molecules are considered (\ch{PhCN} and \ch{PhNO2}) to assess the ability of DFTB to probe a wider range of Raman characteristics, while retaining a quantitative description of the experimental signals.

\begin{figure*}[h]
	\centering
	\includegraphics[width=.99\linewidth]{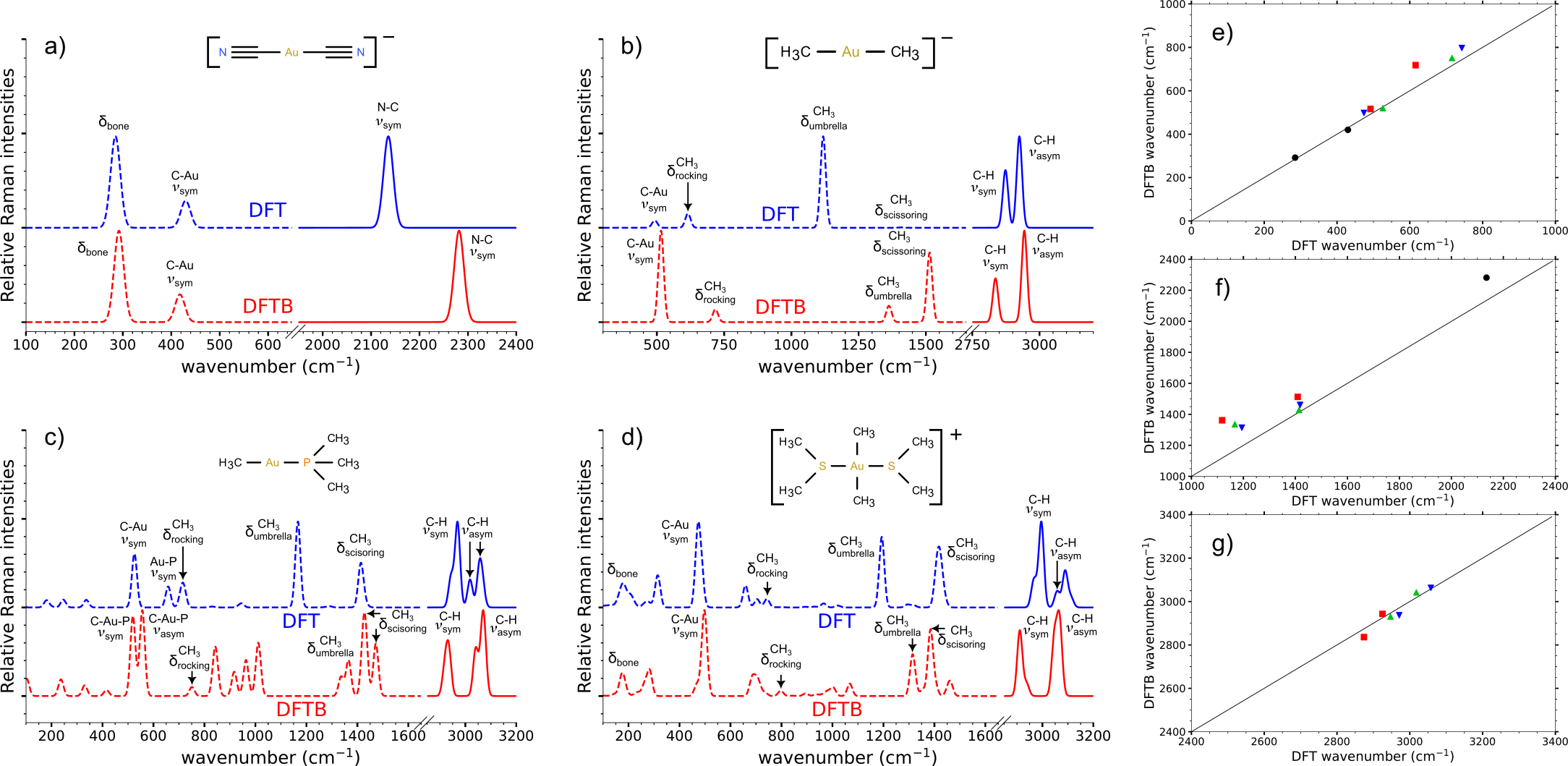}
	\caption{%
		\textbf{a} Simulated Raman spectra of \ch{[Au(CN)2]-} (\textbf{a}), \ch{[AuMe2]-} (\textbf{b}), \ch{[MeAuPMe3]} (\textbf{c}) and  \ch{[Me2Au(Me2S)2]+} (\textbf{d}). Red and blue lines corresponds to spectra computed with DFTB and DFT, respectively. The spectra are fitted with Gaussian functions, and the intensities are rescaled in each frequency window.
		\textbf{e-g} Correlation between the DFT and DFTB Raman-active modes. Black, red, green, and blue symbols represents modes of \ch{[Au(CN)2]-}, \ch{[AuMe2]-}, \ch{[MeAuPMe3]}, and \ch{[Me2Au(Me2S)2]+}, respectively.
	}
	\label{fig:complexes}
\end{figure*}

\subsection{Raman spectra of gold complexes}

To assess the accuracy of DFTB for the Raman response of molecular on gold-carbon bonds, we selected four gold complexes for which experimental data are available based on gold(I) and gold(III) centers, including L- and X-type ligands: \ch{[Au(CN)2]-}, \ch{[AuMe2]-}, \ch{[MeAuPMe3]} and \ch{[Me2Au(Me2S)2]+}. 
Geometry relaxations with both DFT and DFTB lead to the expected linear coordination for gold(I) complexes, and square-planar configuration for the gold(III) complex (see Figure S1,S2, ESI{\dag}). 
Interestingly, the discrepancies on bond lengths are less than 1.39\%, even less so, when considering Au-C bonds only (1.35\%), while angles vary at most by 1.6$^{\circ}$. 
The main differences in bond lengths are caused by the slight DFTB shortening of Au-P (2.26~{\AA} vs. 2.31~{\AA}) and elongation of Au-S (2.44~{\AA} vs. 2.37~{\AA}) as compared to DFT. 
Hence, the DFTB optimized geometries are found always close to the DFT reference, regardless of the nature of the ligand or the oxidation state of the metal. 
This is remarkable given the general purpose of the DFTB parameterization set used (\textit{auorg}) that has been primarily designed for gold bulk/slabs and not for organometallic complexes.

Figure~\ref{fig:complexes} presents the DFT and DFTB simulated Raman spectra for the four gold complexes.
For clarity, different scaling factors are applied on the Raman intensities in low- and high-frequency regions, allowing to avoid one high intensity to invisible the low intensities. Each region are rescaled so the highest peak intensity in the region is 1. Raman spectra with a unique scaling factor on the whole frequency range can be found in ESI{\dag} together with the representations of active normal modes (Figures S7-S14).
The most intense signals in the computed spectra are labeled according to the nature of the active mode: bond stretching ($\nu$), deformation of the main structure of the molecule ($\delta_{\text{bone}}$), methyl rocking ($\delta^{\ch{CH3}}_{\text{rocking}}$), scissoring ($\delta^{\ch{CH3}}_{\text{scissoring}}$) and umbrella ($\delta^{\ch{CH3}}_{\text{umbrella}}$).

DFTB computed Raman spectra show an overall satisfying agreement with the DFT spectra throughout the series of complexes. For all compounds, DFTB mode frequencies follow remarkably well the DFT values for stretching modes, with an average relative error of 3.9\%, which becomes slightly lower when considering specifically Au-C modes (3.5\%). 
C-H stretching modes are also very well described compared to DFT with an average error less than 0.8\%. 
The largest peak displacement among stretching modes is for the $\nu_{\text{N-C}}$ bond in \ch{[Au(CN)2]-} (6.9\%), with a DFTB frequency (2282~cm$^{-1}$) overestimated by 147~cm$^{-1}$ compared  to the DFT value (2135~cm$^{-1}$). 
In the case of \ch{[MeAuPMe3]},  $\nu_{\text{Au-C}}$ is correctly described, while the Au-P stretching mode is underestimated by \textit{ca.} 100~cm$^{-1}$. 
A Similar behavior is observed for \ch{[Me2Au(Me2S)2]+}: $\nu_{\text{Au-S}}$ is obtained with a deviation to DFT twice as large compared to the $\nu_{\text{Au-C}}$.
This overall accuracy is noticeable, and shows that the design of the \textit{auorg} parameters, more specifically the repulsive potentials, by using a straightforward construction (Au-X stretching profile)~\cite{Fihey2015a} intrinsically carry the necessary ingredients to reproduce the vibration frequencies of the corresponding bonds in molecular complexes, even without a specific fine-tuning for vibration frequencies that had been applied for other sets.\cite{3ob-freq} On the contrary, for $\nu_{\text{Au-P}}$ and $\nu_{\text{Au-S}}$, the parameterization of the repulsive potential was conducted using small gold clusters, encompassing thus the effects of neighboring gold atoms as expected for instance in functionalized gold materials, but deviating from a pure Au-X situation.\cite{Fihey2015a,Vuong2020}
The relative intensities of the DFT stretching peaks are also globally well reproduced  with DFTB, except for the asymmetric stretching C-H modes that appear overestimated.
Importantly, DFTB results and reported experimental values (when available) are also in close agreement, with deviations smaller than 3.5\% for all complexes (see Table S3 for reported experimental frequencies), expect for \ch{[Me2Au(Me2S)2]+} where the error raises up to 10.3\% for $\nu_{\text{Au-C}}$, though in this case DFTB values are closer to the measured one than DFT.~\cite{Shaw1973, Rice1975, Bruni1993, Scovell1970} 

The Raman frequencies and intensities of the ligands bone deformation modes for \ch{[Au(CN)2]-} and \ch{[Me2Au(Me2S)2]+} are obtained in excellent agreement with respect to DFT, resulting in errors of 2.5\% and 1.1\%, respectively. 
The situation is more contrasted in the description of \ch{CH3} rocking modes with relatively modest errors for \ch{[MeAuPMe3]-} and \ch{[Me2Au(SMe2)2]+} (4.9\% and 7.3\%, respectively), but significant shift of the mode for \ch{[AuMe2]-} leading to an error of 16.6\%. 
\ch{CH3} scissoring are also found challenging to reproduce, with an error in the peak position that can reach 7.4\% for \ch{[AuMe2]-} and splittings observed for \ch{[MeAuPMe3]-} and \ch{[Me2Au(SMe2)2]+} in DFTB spectra, that are not observed in the DFT results, although the DFT and DFTB mode frequencies are in good agreement (error of 1.0\% and 3.0\%, respectively).
The main discrepancy is obtained on the description of \ch{CH3} umbrella modes with large errors compared to DFT for \ch{[AuMe2]-}, \ch{[MeAuPMe3]}, and \ch{[Me2Au(Me2S)2]+} (21.8\%, 14.5\%, and 10.1\%). Comparison with experiment is made difficult by uncertainty in the determination of the nature of the \ch{CH3} modes. 
However, Raman signals are observed for \ch{[MeAuPMe3]} between 1159 and 1170~cm$^{-1}$,~\cite{Shaw1973, Rice1975} which corresponds to the umbrella modes described by DFT, and confirms the problematic description of those modes by DFTB. 
Contrarily to the above described stretching modes, these deformations are not directly followed in the repulsive potentials, and are intrinsically more sensitive to the pair based tight-binding approximations.

For simulations of Raman spectroscopy, it has been proposed in previous works to quantify the deviations from the reference within separated frequency regions: (i) a global one for all frequencies, (ii) a low-frequency one, below 1000~cm$^{-1}$, (iii) a mid-frequency one for 1000-2000~cm$^{-1}$ range, and (iv) a high frequency one, above $\geq$2000~cm$^{-1}$.~\cite{Zapatatrujillo2023}
By doing so, the median true error for GGA functionals was found close to 60~cm$^{-1}$ for the global frequency range, and 45~cm$^{-1}$ for specific frequency regions.~\cite{Zapatatrujillo2023} 
Inspired by this error analysis, we decompose the deviation between DFTB and DFT peak positions into low-, middle-, and high-energy windows in Figure~\ref{fig:complexes}e-g.  By doing, so we observe a striking concordance between both approaches in the low- and high-energy regimes.
However, this agreement deteriorates in the mid-range, where the scissoring and umbrella modes are observed. Within our DFTB/DFT comparison, the mean deviation attains 57~cm$^{-1}$ over all frequencies, and decreases to 36 and 40 for low and high frequencies, respectively, while it attains 116~cm$^{-1}$ in the mid-frequency zone. The mean absolute relative error among all modes considered is 5.9\% compared to DFT and 6.5\%, 9.6\% and 1.6\% when analysing separetely the low, mid and high frequency regions respectively.

The systematic comparison of DFT and DFTB computed Raman spectra on those four complexes, shows that (i) the qualitative agreement is obtained regardless of the ligands or of the gold oxidation state, (ii) despite, fairly large errors on specific modes (\ch{CH3} scissoring and umbrella modes), excellent accuracy is delivered by DFTB on the position and intensity of Raman peaks, especially in the low- and high-energy regions.
These results comfort us in the use of DFTB to simulate Raman response of molecules on a gold support, carried in the next section, especially to probe the Au-C stretching bonds, often tracked experimentally.

\subsection{On the choice of gold substrate representation}

On the one hand, functionalized gold surface have often been modeled using moderate-sized molecular cluster, as DFT periodic simulations of Raman spectra remain scarce and limited to small unit cells.~\cite{carb2020}
Among those molecular approaches, \ch{Au20} cluster has been widely used to model the interactions between organic compounds and a gold substrate (surface or nanoparticle),~\cite{Laurentius2011, Ahmad2014,Berisha2018,Mohamed2018,Shen2019,Luo2020} thanks to its small size,  its experimental existence, and its geometry encompassing different environment: a planar facet, an edge, or a single atom tip, that can be used to model the non-planar environment of a gold nano-object.
On the other hand, the use of periodic slabs allow the description of functionalized surfaces, or the interface between gold nano-objects and organic molecules from a collection of interfaces between the molecules and various crystalline orientations. Here, we consider the (111) orientation, which is the most stable one for gold (see Figure S5 for details on the slab model).
In the following we compare the evolution of the Raman signature of a simple gold-\ch{CH3} system obtained using these different bonding situations on \ch{Au20}, or using a periodic DFTB calculation on a functionalized gold slab. For comparison purposes, we also include the results obtained on the smallest gold organic model, namely a \ch{AuCH3} molecule, as well as experimental results obtained for a \ch{(CH2)4-COOH} molecule grafted on a gold surface presenting, a similar  C (\textit{sp$^3$})-Au surface interaction.~\cite{Berisha2018}

\begin{figure}[ht]
	\centering
	\includegraphics[width=.75\linewidth]{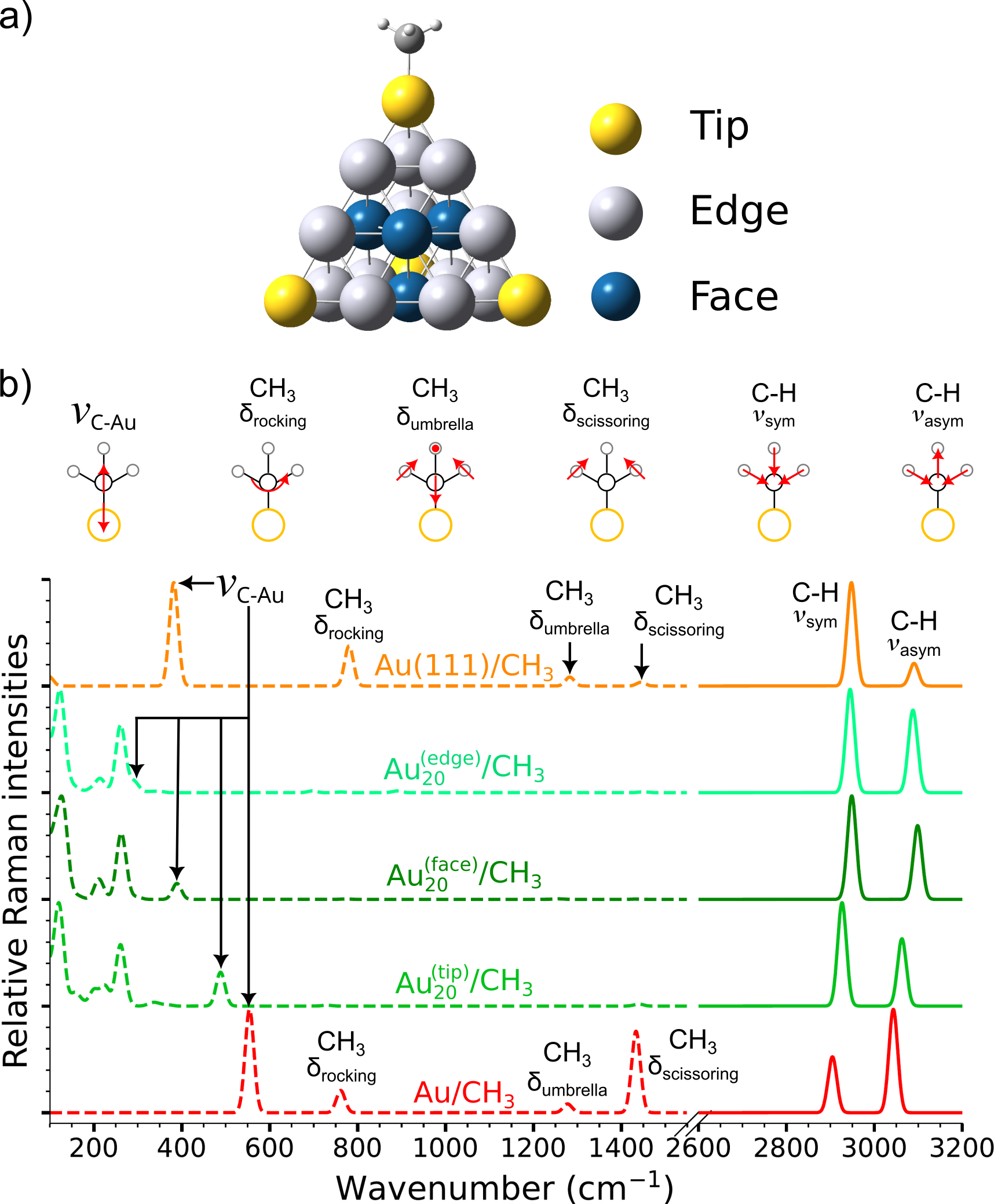}
	\caption{%
		\textbf{a} Representation of \ch{CH3} grafted on a tip site of \ch{Au20}. Gold, carbon and hydrogen atoms are represented in yellow, gray, and white, respectively.
		\textbf{b} DFTB computed Raman spectra of \ch{CH3} grafted on the different sites of a \ch{Au20} cluster.
		The spectra are fitted with Gaussian functions.
		In all the spectra, the most intense signal is set to 1. For clarity, different scaling factor are applied on the Raman intensities in low- and high-frequency regions, allowing to avoid one high intensity to invisible the low intensities. Each region are rescaled so the highest peak intensity in the region is 1.
	}
	\label{fig:au20}
\end{figure}

The different adsorption sites on \ch{Au20} (tip, edge, and face) are represented on Figure~\ref{fig:au20}a. The DFTB adsorption energies of \ch{CH3} shows the clear preference for the tip site ($-$1.69~eV) over the edge ($-$1.21~eV) and face $-$0.94~eV) sites. This is caused by the higher reactivity of the under-coordinated gold atoms found at the edge and the tip, more likely to create a bond with the organic compound.
Corresponding DFTB relaxed geometries are detailed in ESI{\dag} (Figure S4).
The Au-C distances vary from 2.12 to 2.19 {\AA} depending on the various anchoring situations, while the shortest bond distance is for the methyl grafted on a single gold atom, \ch{AuCH3} with a value of 2.04 {\AA}.
The relaxed geometry of the methyl on the Au(111) surface (Figure S6), leading to an adsorption energy of $-$1.98~eV, is close to the one obtained by grafting \ch{CH3} on the face of a \ch{Au20} cluster, with both configuration adsorbed on top site, and a Au-C bond length of 2.19~{\AA}, being identical.

\begin{table*}[ht]
	\centering
	\begin{tabular}{lcccccc}
	\hline													
	&	\ch{AuCH3}	&	\ch{Au20^{face}}\ch{CH3}	&	\ch{Au20^{edge}}\ch{CH3}	&	\ch{Au20^{tip}}\ch{CH3}	&	Au(111)/\ch{CH3}	&	Exp.~\cite{Berisha2018}		\\
	\hline													
$\nu_{\mathrm{Au-C}}$	&	555	&	390	&	293	&	488	&	382	&	387		\\
$\delta_{\mathrm{rocking}}$	&	762	&	768	&	697, 761, 888	&	733	&	780	&			\\
$\delta_{\mathrm{umbrella}}$	&	1276	&	1252	&	1301	&	1278	&	1282	&			\\
$\delta_{\mathrm{scissoring}}$	&	1434	&	1432	&	1450	&	1439	&	1445	&			\\
$\nu^{\mathrm{sym}}_{\text{C-H}}$	&	2905	&	2947	&	2945	&	2926	&	2948	&			\\
$\nu^{\mathrm{asym}}_{\text{C-H}}$	&	3044	&	3099	&	3088	&	3062	&	3090	&			\\
	\hline																		
	\end{tabular}
	\caption{DFTB computed energy of normal modes (cm$^{-1}$) for \ch{CH3} grafted on a single Au atom, on the face, the edge, or the tip of a \ch{Au20} cluster, on a Au(111) surface. The experimental value is obtained for the valeric group (\ch{(CH2)4-COOH}) grafted on a gold surface.~\cite{Berisha2018}}
	\label{tab:ch3}
\end{table*}

The impact of the choice of the model and of the anchoring site on the Raman signal is highlighted on Figure~\ref{fig:au20}b and Table~\ref{tab:ch3}. 
As expected, \ch{CH3} modes ($\delta_{\mathrm{rocking}}$, $\delta_{\mathrm{umbrella}}$, $\delta_{\mathrm{scissoring}}$, $\nu_{\text{C-H}}$), which do not directly involve gold atoms, are nearly not impacted by the model chosen.
Still, we observe that for \ch{Au20^{edge}}\ch{CH3}, the $\delta_{\mathrm{rocking}}$ mode is split in 3 due to the low symmetry of this adsorption site.
The situation is different for the stretching mode $\nu_{\text{Au-C}}$. 
The most simple model, the single Au atom, returns a frequency very far the experimental value obtained for gold surface,~\cite{Berisha2018} with a 40\% deviation. 
Such discrepancy is not expected for a type of mode well-described by DFTB (\textit{vide supra}), and is instead more representative of the very different chemical environment in this trivial molecular model.
This deviation is halved when considering the methyl grafted at the tip of \ch{Au20} (\ch{Au20^{tip}}\ch{CH3}) but remains extremely high. 
An excellent agreement on the position of the $\nu_{\text{Au-C}}$ mode is obtained when the methyl is placed on one of its face, with a deviation of less than 2\%. This surface-like adsorption site is clearly able to mimic the extended environment of the experimental gold substrate for the Raman profiles.
Note that for \ch{Au20} cluster, several broad peaks with high intensities appear bellow 400~cm$^{-1}$. Those peaks corresponds to vibrational modes of the gold cluster.

The DFTB simulated spectra for \ch{CH3} on the periodic Au(111) surface (Figure~\ref{fig:au111}b) presents similarities with the one obtained on the face of \ch{Au20}. 
In particular, the stretching mode $\nu_{\text{Au-C}}$ (382~cm$^{-1}$) is described with a deviation of 1.3\% with respect to the experimental value of 387~cm$^{-1}$, confirming the quality of description of DFTB for this type of mode. 
As previously observed, the position of the $\delta_{\mathrm{rocking}}$, $\delta_{\mathrm{umbrella}}$, $\delta_{\mathrm{scissoring}}$, and $\nu_{\text{C-H}}$ modes, that do not involve the Au-C bond, is not sensitive to the model used and similar values are obtained.
The intensities, however, varies and the Au(111) approach allows the recovering of an intense band for $\nu_{\text{Au-C}}$, as well as a more intense peak for the symmetric stretching mode $\nu_{\text{Au-C}}^{\text{sym}}$ with respect to $\nu_{\text{Au-C}}^{\text{a-sym}}$, a feature also observed experimentally on gold complexes.~\cite{Shaw1973}

Both \ch{Au20^{face}} model and a full periodic calculation in DFTB are thus able to reproduce the key experimental Raman signatures of a \textit{sp$^3$} carbon derivative on a gold surface. 
Other \ch{Au20} models (edge, tip) are less adapted for this specific case, but may be able to grasp the specific interactions on the non-planar environment of gold nano-objects.

\subsection{Common functionalization of gold surfaces}

\begin{figure}[ht]
	\centering
	\includegraphics[width=.75\linewidth]{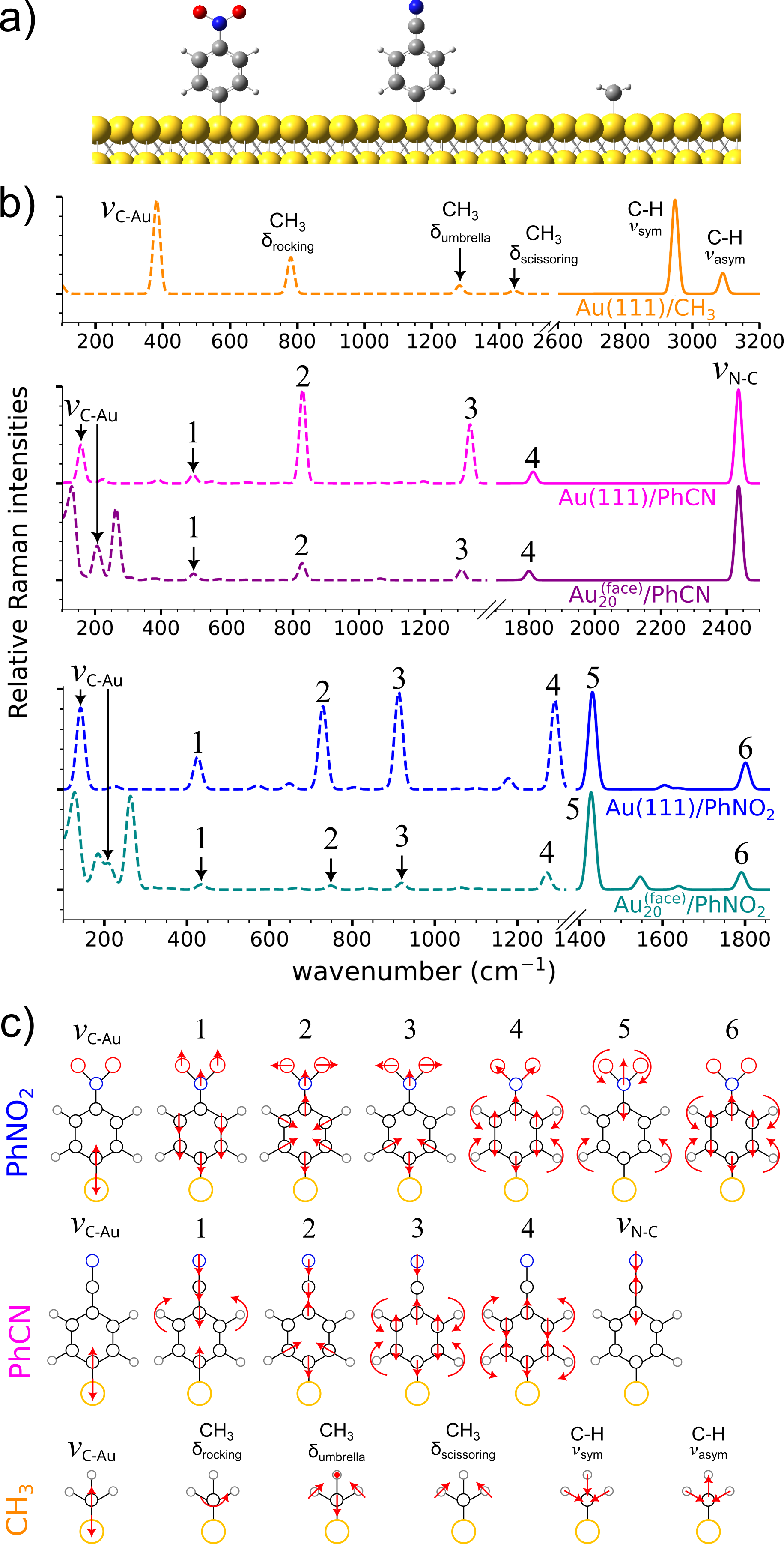}
	\caption{%
		\textbf{a} \ch{CH3}, \ch{PhNO2}, and \ch{PhCN} grafted on the Au(111) surface. Gold, oxygen, nitrogen, carbon and hydrogen atoms are represented in yellow, red, blue, gray, and white, respectively.
		\textbf{b} Corresponding DFTB computed Raman spectra.
		The spectra are fitted with Gaussian functions.
		For clarity, different scaling factor are applied on the Raman intensities in low- and high-frequency regions, allowing to avoid one high intensity to invisible the low intensities. Each region are rescaled so the highest peak intensity in the region is 1.
		\textbf{c} Schematic representation of the highlighted modes for Au(111)/\ch{PhCN} and Au(111)/\ch{PhNO2}.
	}
	\label{fig:au111}
\end{figure}

 In the following we shift our focus on more complex chemical groups often used for gold surface functionalization trgough a Au-C bond, namely cyanophenyl (\ch{PhCN}) and nitrophenyl (\ch{PhNO2}), (Figure~\ref{fig:au111}a).~\cite{Laurentius2011,Ahmad2014,Mohamed2018,Luo2020} 
 In each case, we details the DFTB geometries and Raman spectra, focusing principally on the description of the stretching modes, often tracked experimentally to probe the functionalization of the support.

The DFTB geometry optimization of \ch{PhCN} and \ch{PhNO2} on \ch{Au20^{face}} and Au(111) leads to similar geometry characteristics (Figures S4 and S6) where the molecules are binding vertically on the surface on a top site. In particular, the Au-C bond lengths are nearly identical (2.22~{\AA} and 2.23~{\AA}, respectively). 
DFTB computed Raman spectra for Au(111)/\ch{PhCN} and Au(111)/\ch{PhNO2} show a greater complexity than the methyl case, due to the coupling with the numerous modes of the phenyl ring (Figure~\ref{fig:au111}b) that can be traced back to the expected in deformations in the D$_{6h}$ point group.\cite{Preuss2006}
Mode frequencies are reported in Tables~\ref{tab:au111-phcn}.
A schematic representation of the modes is given in Figure~\ref{fig:au111}c.

In addition to the stretching modes $\nu_{\text{Au-C}}$ and $\nu_{\text{N-C}}$, we observe four active mode involving the phenyl for \ch{PhCN} on Au(111). 
Mode 1 is dominantly composed of the \textit{e$_{2g}$} mode of D$_{6h}$ benzene (in-plane deformation) combined with the the Au-C stretching. 
Mode 2 presents the same coupling with the \textit{b$_{2u}$} in-plane deformation of the phenyl ring. 
Mode 3 corresponds to the \textit{a$_{1g}$} mode (phenyl breathing), similar to mode 4, close to the \textit{e$_{2g}$} mode of benzene.
A similar description can be carried out for \ch{PhNO2} on Au(111), where mode 1 is found to mix the \textit{e$_{2g}$} mode of benzene (in-plane deformation), \ch{NO2} stretching, and $\nu_{\text{Au-C}}$ mode. 
mode 2 and 3 correspond to a \textit{b$_{2u}$} mode of benzene coupled to Au-C stretching, and \ch{NO2} bending modes. 
mode 4 is the \textit{a$_{1g}$} phenyl breathing coupled with with N-O stretching. 
mode 5 couples the \textit{e$_{1u}$} mode of benzene with \ch{NO2} scissoring. 
Finally, mode 6 exhibit the same features as the benzene \textit{e$_{2g}$} mode.
One should note that mode 2 in the \ch{PhCN} case and modes 2 and 3 in the \ch{PhNO2} case present the same phenyl deformation. 
A match can be also be found between mode 3 and 4 of \ch{PhCN} and mode 4 and 6 of \ch{PhNO2}.

Some of these modes have been assigned on the experimental spectra allowing for a comparison with our results (Tables~\ref{tab:au111-phcn}).~\cite{Mohamed2018,Luo2020}
Among those, the breathing mode of the phenyl ring of Au(111)/\ch{PhNO2} (mode 4) is found with a 2.9\% deviation with respect to the experimental value reported on gold nanoparticles with a diameter of 15$\pm$3~nm.~\cite{Luo2020}
The discrepancy is larger for the corresponding mode in the \ch{PhCN} system (mode 3), with a deviation reaching 15.3\% , using data from the same set of measures.
This further increases, up to a very large 74.1\% when using the labeling of the modes from another set of measures obtained on nanoparticles presenting a diameter smaller than 5~nm, and for which the Au(111) approximation might be less adequate.~\cite{Mohamed2018}
The N-C stretching $\nu_{\text{N-C}}$ is fairly reproduced with an error of 9.4\%.
Overall, the computed modes frequencies are close when using either Au(111) or \ch{Au20} models for \ch{PhCN} and \ch{PhNO2}, but intensities show striking differences, as already observed for \ch{Au20^{face}CH3}.

\begin{table}[ht]
	\centering
    \small
	\begin{tabular}{lccccc}
	\hline						
	 &	\ch{Au20^{face}}\ch{PhCN}	&	Au(111)/\ch{PhCN}	&	Exp.	\\
	\hline						
$\nu_{\mathrm{Au-C}}$	&	206	&	157	&	488~\cite{Mohamed2018}, 400~\cite{Luo2020}	\\
Mode 1	&	499	&	497	&		\\
Mode 2	&	828	&	829	&		\\
Mode 3	&	1311	&	1337	&	768~\cite{Mohamed2018}, 1580–1595~\cite{Luo2020} 	\\
Mode 4	&	1799	&	1812	&		\\
$\nu_{\mathrm{N-C}}$	&	2436	&	2435	&	2227~\cite{Mohamed2018}, 2220~\cite{Luo2020}	\\
	\hline						
	 &	\ch{Au20^{face}}\ch{PhNO2}	&	Au(111)/\ch{PhNO2}	&	Exp.\\	
	\hline						
$\nu_{\mathrm{Au-C}}$	&	210	&	142	&	400~\cite{Luo2020}	\\
Mode 1	&	435	&	425	&		\\
Mode 2	&	749	&	729	&		\\
Mode 3	&	920	&	912	&		\\
Mode 4	&	1271	&	1291	&	1330~\cite{Luo2020}	\\
Mode 5	&	1428	&	1430	&		\\
Mode 6	&	1791	&	1798	&		\\
	\hline				
	\end{tabular}
	\caption{Experimental and DFTB computed energy of normal modes (cm$^{-1}$) highlighted on Figure~\ref{fig:au111}b for \ch{PhCN} and \ch{PhNO2} grafted on Au(111). The assignment of the experimental active modes is deduced from the references cited in the table and discussed in the text.}
	\label{tab:au111-phcn}
\end{table}

						

Experimentally, the Au-C stretching modes for \ch{PhCN} and \ch{PhNO2} on Au(111) are assigned to peaks around 400~cm$^{-1}$,\cite{Laurentius2011} while DFTB stretching modes are found at 157 and 142~cm$^{-1}$, respectively. Such a discrepancy is clearly out of the previously observed trends, where stretching modes are expected to be well described by DFTB.
In this case, it is critical to note that the experimental assignment of Au-C stretching mode is based on a comparison with the expected  frequency in molecular gold complexes, thus introducing potential biases.~\cite{Socrates2004, Shaw1973, Bruni1993, Laurentius2011, Ahmad2014, Luo2020}
As established here, and in a previous work,~\cite{Tetenoire2024} the choice of the model for the gold substrate (single atom, ion, small cluster, or crystalline surface) greatly affects the position of the $\nu_{\text{Au-C}}$ mode. Computationally, it is also easy to mistake the mode 1 of Au(111)/\ch{PhCN} or Au(111)/\ch{PhNO2} for the stretching mode because the distortion of the phenyl induces a variation of the Au-C distance and therefore mimics the stretching mode.
For these two systems, the experimental spectra do not range from a sufficiently low energy to confirm the presence of a Raman active mode around 150~cm$^{-1}$.
However, this hypothesis has been verified recently in the case of a macrocycle grafted through phenyl groups on a gold nanorod, for which the stretching modes have been identified around 250~cm$^{-1}$.~\cite{Tetenoire2024}
In addition we also show a clear impact of the nature of the organic moiety on   $\nu_{\text{Au-C}}$, moving from 382~cm$^{-1}$ for a methyl, to 218~cm$^{-1}$ for a phenyl (previous work),~\cite{Tetenoire2024} to bands shifted to lower frequencies when a substituent of an increasing electron withdrawing character is added to the phenyl: 157~cm$^{-1}$ for {\ch{PhCN}}, and 142~cm$^{-1}$  for {\ch{PhNO2}}.

\section{Conclusion}

This work evaluated the performance of Density Functional Tight Binding (DFTB) for simulating Raman spectra of molecular gold complexes and small organic molecules (namely \ch{CH3}, \ch{PhCN} and \ch{PhNO2}) grafted on a gold support (\ch{Au20} and Au(111) surface) by comparing to reference DFT and experimental data. For all systems, our analysis reveals that DFTB achieves excellent agreement with DFT and experimental Raman profiles in the low-frequency region (200–700~cm$^{-1}$), particularly for Au-C stretching modes. However, discrepancies emerge in the mid-frequency region (700–2000~cm$^{-1}$), where DFTB struggles to accurately model for instance \ch{CH3} rocking, umbrella, and scissoring modes, highlighting the current limitations in the tight-binding approach when simulating ligand vibrations that involve more than a single bond.
In contrast, the high-frequency region (2000–3000~cm$^{-1}$) shows generally excellent agreement, as this regions encompass for instance C-H and C$\equiv$N stretching modes.

For organic molecules grafted on a gold substrate, the importance of choice of the model to represent the metallic region is highlighted. The popular \ch{Au20} cluster proved to be a satisfying approximation of a extended gold surface, only when the molecule is grafted onto a face of the cluster, while the edge and the tip of \ch{Au20} induce very different Raman profiles.
However, in any case, using \ch{Au20} leads to a mismatch of the intensity profiles obtained on Au(111) surfaces. 
The scalability of DFTB enables tractable simulations of extended systems, circumventing the limitations of cluster models and providing a more realistic representation of interfacial bonding, ultimately allowing the theoretical identification of a primary Au-C stretching frequency for phenyl-based molecules on Au(111) near 200~cm$^{-1}$, thus contrasting with the frequently found experimental identifications based on the extrapolation of Au-C stretching frequency from molecular complexes to the gold material. 

Taking advantage of this fast computational protocol, further following works will tackle the dependency on the shape and size of the nano-object, \textit{i.e.} beyond the Au(111) model, hence offering a more complete description of the complex experimental system. Of course, refining DFTB parametrization for mid-frequency modes is clearly another lead for improving accuracy in modeling ligand deformations and reaching a full quantitative accuracy. 
By showing the success and limitations of DFTB method, this work establishes a new transformative approach for simulating Raman spectra of gold-organic interfaces, bridging quantum-scale precision with nanoscale applicability.

\section*{Conflicts of interest}

There are no conflicts to declare.

\section*{Data availability}

The code to produce the Raman spectra from DFTB+ and ADF data and the outputs from DFTB+ and ADF calculations are available at the following open repository: https://doi.org/10.5281/zenodo.20639302.

\section*{Acknowledgements}

This work was supported by the ANR MARCEL project, grant ANR-21-CE50-0034-MARCEL, of the French Agence Nationale de la Recherche. The authors are thankful to Dr. Karine Costuas for the fruitful discussions on DFT protocols for Raman spectra simulations.

\balance


\bibliography{bibliography} 
\bibliographystyle{rsc} 

\end{document}